# FUNDAMENTALITY OF THE SOMMERFELD'S FINE STRUCTURE CONSTANT IN BRIDGE THEORY AND CONSEQUENCES ON THE FUNDAMENTAL ATOMIC CONSTANTS

Massimo Auci[1], Giuseppe Basile[2], Ugo Fabbri[3]

[1] OdisseoSpace. Space Science dept.
Via Battistotti Sassi 13 – 20133 Milano, Italy
[2] INRiM. Istituto Nazionale di Ricerca Metrologica
Strada delle Cacce 91 -10135 Torino, Italy
[3] Via Conti 19 – 34100 Trieste, Italy

The Bridge Theory (BT) applied to a simple hydrogen atom model proves that the current values for the electromagnetical physical constants could be affected by a small error due to a non completely corrected modelling of the physical phenomenology.

## Introduction

The BT [1-2-3-4] develops from the consideration that the effective spatial symmetry that characterises an electromagnetic (em) wave depends on the nature of the source. If we consider an ideal point-like source of em waves, the propagation occurs radially with spherical wave fronts, so that also the Poynting vector (PV) will be radial. We shall call "ideal" such a source. On the other hand, in nature one never deals with "ideal" sources. The simplest source that can be produced is at least endowed with a dipole moment, and therefore cannot be considered point-like. As a consequence, the propagation will not occur by a spherical wave front and the PV will not be radial everywhere. Hence, the PV will have a non-zero transverse component. We shall call "real" such a source.

Usually we assume that at a distance from the source much greater than the emission wavelength, the wave has total spherical symmetry. This is not true and at short distances, the non-spherical symmetry introduces non-negligible approximations from the energy point of view of the observer. In this case, the space-time distribution of the non-zero transverse component of the PV reduces the radial emission of energy and, consequently, localises energy around the source that manifests itself as an energy grain with the characteristics of a local virtual photon exchanged by the interacting pair charges.

From the metrological point of view one of the most relevant results of BT concerns the electromagnetical nature of the Sommerfeld's fine structure and Planck constants. In fact, a detailed theoretical analysis of the em field structure near the source zone of the source puts in evidence the dependence of the constants $\alpha$ and $\hbar$ on the internal em interaction parameters, correlating the origin of the Planck's constant with that of the fine structure constant and not vice versa. BT proves that $\alpha$ is fundamental because independent by the value of other physical constants. The Planck constant $\hbar$, on the contrary, needs not to be assumed as fundamental since it depends on the set of values $(\alpha, e, c)$. The dependence of $\hbar$ from $\alpha$ implies that both constants show the same two internal degrees of freedom. The first is associated with the dipole moment length of the interacting pair originating the source, the second with the em field torsion produced by the delay effect in the field propagation. This latter parameter depends on the former and has the role of a structure parameter of the two constants.

These external dependencies suggest neither constants $\alpha$ and $\hbar$ can be considered absolutely and universally as such, because it may happen that in a different physical context with respect to that in which the source is usually produced, they might acquire different values [4].

Assuming as true the theoretical value of the fine structure constant corrected for the angular spread of the interacting charge particle as in ref. [4], its value is exactly coinciding with the experimental one [5] only as a consequence of the value of the angular correction applied. In fact, different angular corrections give very little variation of the value of $\alpha$, without modifying the phenomenology describing the nature of the constants $\alpha$ and $\hbar$. The best values without angular correction is

$$\alpha = \left(\frac{4\pi}{3}F_t + \frac{1}{4\pi\bar{\rho}^2}\right) = 1/137.036(6) \qquad (1)$$

$$\hbar = \left(\frac{4\pi}{3}F_t + \frac{1}{4\pi\bar{\rho}^2}\right)^{-1} \frac{e^2}{c} Js \ , \qquad (2)$$

where

$$F_t = \int_0^\pi \Theta_t (\bar{\rho}, \bar{\theta}) \, d\theta \cong 32.7034(18) \tag{3}$$

is the em structure factor defined by the interaction parameters

$$\bar{\theta} \cong 0.25000(69)\pi \ rad \tag{4}$$
$$\bar{\rho} \cong 1.27555(78) \tag{5}$$

of the source. The value of $\alpha$ depends on the accuracy of the theoretical values (4) and (5).

## Consequences of a stationary null-source in BT

Using BT, we consider the case of a non-emitting real source where the radial component of the PV is everywhere null. We define this interaction a null-source. In the null-source the pair ($e^-e^+$) interacts forming a stable state where reciprocally each of the two charges is captured by the other. The em field propagation occurs only transversally respect to the radial direction of emission and both the interacting charges perfom a "zitterbewegung" motion along a close path, each one around the other. With respect to the lab-frame the motion occurs around the centre of the null-source. The energy produced during the interaction is confined on the spherical surface $\Sigma_0$ supplying the two moving charges.

From the viewpoint of the positive charge, that represents the charge of the nucleus, the negative one performs an orbital path around the nucleus with radius equal to $\lambda_0$. This suggests some considerations. Energy and momentum conservation requires that the total emission is supplied by the energy localized inside the source zone during the interaction. In the null-source there is not emission, hence energy and momentum are free to supply the energy and momentum of (to?) the orbital electron. Since the localised energy agrees with one of a photon having the wavelength of the null-source, the total energy associated to the orbiting electron is

$$E = \frac{2\pi\hbar c}{\lambda_0}. \tag{6}$$

Considering that in the lab-frame the conservation of the angular momentum needs that the energy of emission is balanced by the energy associated to a field spin, in a null-source the field spin is responsible for a spin energy of the two charges. In the nucleus-frame we write

$$E_{spin} = P_t c = \int_0^{2\pi} d\varphi \int_\infty^{\lambda_0} \frac{d}{dt} \vec{p} \times d\vec{r} = 2\pi \frac{e^2}{\lambda_0} \tag{7}$$

eqs. (6) an (7) give the relativistic $\beta$ factor of the orbiting electron

$$\beta = E_{spin}/E = e^2/\hbar c \equiv \alpha. \tag{8}$$

Eq. (8) implies that the negative charge has a fixed orbital velocity and $\gamma$ Lorentz's factor, with values independent of the effective shape of the orbit, i.e. of the energy of the null-source:

$$\gamma = 1/\sqrt{1-\alpha^2} \cong 1.0000266. \tag{9}$$

Considering the rest mass of an electron as in ref. [5]

$$E = m_e c^2 = 0.51099906(15) \ \text{Mev}, \tag{10}$$

the eqs. (1) and (6) give

$$\lambdabar_0 = \left(\frac{e^2}{m_e c^2}\right) \alpha^{-1} \tag{11}$$

where, in agreement with [5], the ratio in brackets is the classical electron radius $r_e$. The eq. (11) corresponds to the standard value of the Compton wavelength of the orbiting electron.
Applying the transformation [6]

$$\lambda_0 = \beta\gamma\lambda = \lambda\alpha\gamma \tag{12}$$

to the Compton wavelength (11), using eq. (8) we obtain

$$\lambdabar = r_e \alpha^{-2} \gamma^{-1} \cong 0.5291632 \cdot 10^{-10} \, \text{m} \tag{13}$$

which in the lab-frame is the measure of the mean radius of the orbit on which the electron is moving. The Eq. (13) has the same role of Bohr radius but has a value lower than the standard one of a Lorentz's factor (9). Since in the lab-frame the kinetic energy of the orbiting electron must balance the Rydberg energy $E_I = \hbar c R_\infty$ required for the atom ionization:

$$\hbar c R_\infty = m_e c^2 (\gamma - 1), \tag{14}$$

the right side of eq. (14) gives

$$E_I = m_e c^2 \left( \frac{1}{\sqrt{1-\alpha^2}} - 1 \right) \cong 13.60637 \, \text{eV}. \tag{15}$$

Eq. (15) defines the ionization potential only a bit much larger than the standard value of the Rydberg energy. In fact from the definition in ref. [5] we have

$$E_I = m_e c^2 \alpha^2 / 2 \cong 13.60569 \, \text{eV}, \tag{16}$$

that is lower than that given by eq. (15) because the factor $\alpha^2/2$ appearing in the definition (16) is coinciding only with the first-order term of the Taylor series expansion of the relativistic kinetic factor:

$$\gamma - 1 = \frac{1}{\sqrt{1-\alpha^2}} - 1 = \frac{1}{2}\alpha^2 + \frac{3}{8}\alpha^4 + \ldots \tag{17}$$

In this sense eq. (16) can be considered a non relativistic approximation of eq. (15).

## Conclusion

Starting from the fundamentality of the fine structure constant and from its theoretical evaluation (1), BT seems to be a promising instrument to review and adjust the em phenomenology and constants, thus giving a new, possibly more complete vision of the nature.

## References

[1] M.Auci. "A Conjecture on the physical meaning of the transversal component of the Poynting vector". Phys. Lett. A 135 (1989) 86.
[2] M.Auci. "A Conjecture on the physical meaning of the transversal component of the Poynting vector. II. Bounds of a source zone and formal equivalence between the local energy and photon". Phys. Lett. A 148 (1990) 399.
[3] M.Auci. "A Conjecture on the Physical meaning of the transversal component of the Poynting vector. III. Conjecture proof and physical nature of fine structure constant". Phys. Lett. A 150 (1990) 143.
[4] M.Auci. and Guido Dematteis, "An Approach to Unifying Classical and Quantum Electrodynamics". IJMP B, Vol. 13, No.2 (1999) 1525.
[5] Particle Data Group. Phy. Rev. D50 (1994) 1173
[6] M.Auci. "Special relativity and quantum nature of Matter in Bridge Theory". Internet preprint diffusion., phy.auciwebgate.it